\documentclass[aps,prb,twocolumn,amsmath,amssymb,floatfix,superscriptaddress]{revtex4-1}

\usepackage{amsmath}
\usepackage{amssymb}
\usepackage{bm}
\usepackage{xcolor}
\usepackage[utf8]{inputenc}
\usepackage{graphicx}
\usepackage{dsfont} 
\usepackage{hyperref}

\hypersetup{colorlinks=true,linkcolor=blue,urlcolor=blue,citecolor=blue}

\usepackage{amsfonts}
\usepackage{braket}
\usepackage{dsfont}

\begin{document}

\title{Dispersive Landau levels and valley currents in strained graphene nanoribbons}
\author{\'{E}tienne Lantagne-Hurtubise}
\email{lantagne@phas.ubc.ca}
\affiliation{Department of Physics and Astronomy \& Stewart Blusson Quantum Matter Institute, University of British Columbia, Vancouver, BC, Canada V6T 1Z4}
\affiliation{Kavli Institute for Theoretical Physics, University of California, Santa Barbara, CA 93106, USA}
\author{Xiao-Xiao Zhang}
\email{xiaoxiao.zhang@ubc.ca}
\affiliation{Department of Physics and Astronomy \& Stewart Blusson Quantum Matter Institute, University of British Columbia, Vancouver, BC, Canada V6T 1Z4}
\author{Marcel Franz}
\affiliation{Department of Physics and Astronomy \& Stewart Blusson Quantum Matter Institute, University of British Columbia, Vancouver, BC, Canada V6T 1Z4}

\begin{abstract}
We describe a simple setup generating pure valley currents -- valley transport without charge transport -- in strained graphene nanoribbons with zigzag edges. The crucial ingredient is a uniaxial strain pattern which couples to the low-energy Dirac electrons as a uniform pseudomagnetic field. Remarkably, the resulting pseudo-Landau levels are not flat but disperse linearly from the Dirac points, with an opposite slope in the two valleys. We show how this is a natural consequence of an inhomogeneous Fermi velocity arising in the low-energy theory describing the system, which maps to an exactly-solvable singular Sturm-Liouville problem. The velocity of the valley currents can be controlled by tuning the magnitude of strain and by applying bias voltages across the ribbon. Furthermore, applying an electric field along the ribbon leads to pumping of charge carriers between the two valleys, realizing a valley analog of the chiral anomaly in one spatial dimension. These effects produce unique signatures that can be observed experimentally by performing ordinary charge transport measurements and spectroscopy.
\end{abstract}

\date{\today}

\maketitle

\section{Introduction}
The electronic band structure of graphene~\cite{RevModPhys.81.109} hosts two symmetry-inequivalent Dirac points, leading to an effective pseudospin degree of freedom -- the valley -- at energies close to the charge neutrality point. The possibility of addressing or controlling the valley degree of freedom~\cite{Rycerz2007, Akhmerov2007, Xiao2007, Yao2008} has ushered in the field of ``valleytronics"~\cite{Schaibley2016}, which now extends beyond graphene to gapped 2D Dirac materials~\cite{Lensky2015} such as transition-metal dichalcogenide monolayers~\cite{Xiao2012}, bilayer graphene~\cite{Sui2015, Shimazaki2015,Li2018} and two-dimensional ferroelectrics~\cite{Rodin2016,Chang2019}.

In monolayer graphene, a plethora of valley-specific phenomena has been investigated recently. This includes valley filters and switches~\cite{Rycerz2007, Pereira2009, Zhai2010, Fujita2010, Gunlycke2011, Settnes2016b, Stegmann2018} (which selectively reflect electrons within a given valley), valley beam splitters~\cite{Garcia-Pomar2008,  Settnes2016b, Stegmann2018, Prabhakar2019} (which spatially separate electrons according to their valley index) and waveguides for valley-polarized currents~\cite{Wu2011, Wu2018}. The generation and detection of pure valley currents -- currents transporting only the valley degree of freedom but no charge -- have also received some attention. Theoretical proposals include optical excitations generated by polarized light~\cite{Golub2011}, cyclic strain deformations~\cite{Jiang2013}, quantum pumping~\cite{Wang2014a,Wang2014b} or applying AC bias~\cite{Yu2014}. Valley currents have been observed experimentally as edge states in graphene superlattices~\cite{Gorbachev2014} and graphene bilayers~\cite{Sui2015, Shimazaki2015,Li2018}, but not yet in monolayer graphene. 

The goal of this paper is to describe an alternative way to generate valley currents in monolayer graphene subjected to non-uniform elastic strain.  It is well known that elastic strain can be used to tailor the electronic properties of graphene~\cite{Amorim2016,Si2016,Naumis2017}-- the most dramatic example being the creation of uniform pseudomagnetic fields~\cite{Manes2007, Guinea2009, Low2010, Guinea2010_bending, deJuan2013, Manes2013, Zhu2015, Settnes2016} which lead to Dirac pseudo-Landau levels (pLLs)~\cite{Guinea2009}. This effect was first observed in scanning tunneling microscopy (STM) measurements of graphene ``nanobubbles" grown on a platinum substrate~\cite{Levy2010}.  Subsequent experimental work confirmed this result\cite{Yeh2011,Lu2012,Li2015,Liu2018}, including a recent momentum space observation of pLLs using angle-resolved photoemission (ARPES)~\cite{Nigge2019}. In this work, we show how strain can lead to equilibrium valley currents in graphene nanoribbons through the formation of \emph{dispersive} pLLs.

\begin{figure}
\centering
\includegraphics[width=\columnwidth]{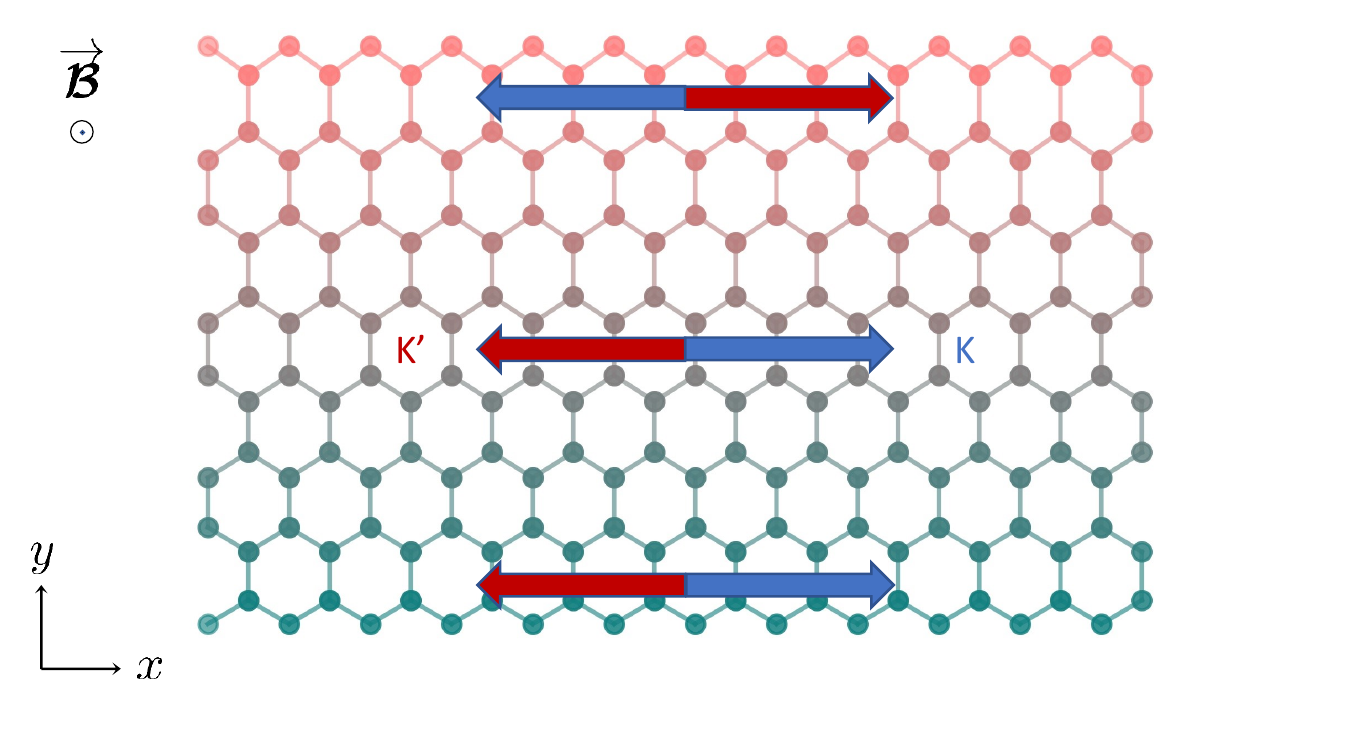}
\caption{Graphene nanoribbon with zigzag edges and periodic boundary conditions along the $x$ direction. An inversion-symmetry breaking uniaxial strain (represented by the color scale) is applied along the $y$ direction and generates a uniform out-of-plane pseudomagnetic field $\bm{\mathcal{B}}$. The low-energy physics of this ribbon is described by dispersing pseudo-Landau levels with an opposite velocity in the two valleys, as shown by blue ($\bm{K}$) and red ($\bm{K}'$) arrows. This leads to equilibrium valley currents in the bulk. Localized edge states also host counter-propagating valley currents. \label{fig:ribbon}}
\end{figure}

Our setup is described in Fig.~\ref{fig:ribbon} -- we consider a graphene nanoribbon with zigzag edges, infinite along the $x$ direction. A static, uniaxial strain pattern is applied along the $y$ direction. The applied strain increases linearly with $y$, generating a uniform pseudomagnetic field $\bm{\mathcal{B}}$ perpendicular to the plane which quantizes the low-energy electronic spectrum to a ladder of pLLs. These pLLs show the $\sim \sqrt{|n|}$ energy spacing characteristic of Dirac electrons but are \emph{not} flat -- instead, they disperse linearly near the Dirac points, with an opposite slope in the two valleys. In equilibrium, the charge current carried by each valley cancels out but the valley current adds up -- thus the bulk of the ribbon carries a net valley current. The edges hosts counter-propagating valley currents (in a direction determined by the sign of $\bm{\mathcal{B}}$) -- thus the ribbon as a whole acts as ``valley-helical" wire (see Fig~\ref{fig:ribbon}). This counter-intuitive feature is naturally understood in terms of an inhomogeneous Fermi velocity~\cite{deJuan2012} arising in this geometry. We provide a novel analytical solution of a singular Sturm-Liouville problem which clarifies this connection. We also discuss how to control the slope of the pLLs, and thus the velocity of the valley currents, by applying bias voltages across the ribbon. To this end, we generalize the seminal solution of Ref.~[\onlinecite{Lukose2007}] to our setup, with Dirac electrons subjected to perpendicular electric and \emph{pseudo}magnetic fields. Finally, we show that applying an electric field along the ribbon leads to pumping of charge carriers between the two valleys, thus realizing an analog to the chiral anomaly in one spatial dimension, and discuss a related \emph{negative strain-resistance} effect observable in electrical conductivity measurements.

The rest of this paper is organized as follows. In Sec.~\ref{Sec:setup} we describe our model and derive the low-energy theory in the presence of uniaxial strain. In Sec.~\ref{Sec:LLsol} we summarize the analytical solution for the bulk pLLs and compare our results to numerical calculations. We also discuss the appearance of valley currents and the chiral anomaly. In Sec.~\ref{sec:scalar_potential} we consider the effects of electric fields (either externally applied or induced by the strain itself) and discuss how they renormalize the slope of the pLLs. We offer concluding remarks in Sec.~\ref{Sec:discussion} and relegate more technical contributions to the Appendices.

\section{The model}\label{Sec:setup}

We consider an infinite graphene nanoribbon (with the periodic direction along $x$), width $W$ and zigzag edges, as shown in Fig.~\ref{fig:ribbon}. We model the system using the nearest-neighbor tight-binding Hamiltonian on the honeycomb lattice, which in absence of strain reads
\begin{align}
H = - t \sum_{<\bm{r}, \bm{r'}>} \left( a_{\bm{r}}^\dagger b_{\bm{r'}} +  b_{\bm{r'}}^\dagger a_{\bm{r}}  \right),
\label{eq:tight_binding}
\end{align}
where $a_{\bm{r}}^\dagger$ ($b_{\bm{r'}}^\dagger$) creates an electron in the $p_z$ orbital on the sublattice A (B), $t = 2.7 \text{eV}$ and the nearest-neighbor distance is $a_0 =0.142$ nm. In view of the negligible spin-orbit coupling in graphene and the absence of (real) magnetic fields in our setup, we neglect the spin degree of freedom in this paper. We incorporate strain into our tight-binding Hamiltonian [Eq. (\ref{eq:tight_binding})] through a simple modulation of the hopping parameters,
\begin{align}
    t \rightarrow t_{\bm{r} \bm{r'}} = t e^{-\gamma \Delta u_{\bm{r} \bm{r'}}},
    \label{eq:hopping_modulation}
\end{align}
where $\Delta u_{\bm{r} \bm{r'}}$ is the displacement of atoms at positions $\bm{r}$ and $\bm{r'}$ relative to $a_0$, and $\gamma = -\partial \ln t/ \partial \ln a |_{a = a_0} \sim 3.37 $ is the Gruneisen parameter of graphene~\cite{Pereira2009b}. We work within the framework of continuum elasticity theory, where the displacement field $\Delta u_{\bm{ r r'}}$ is expressed as a smooth function of the spatial coordinates. This approach is valid for displacement fields varying slowly on the lattice scale. Novel effects can be expected when going beyond the continuum elasticity, see e.g. Ref.~[\onlinecite{Sloan2013},\onlinecite{BarrazaLopez2013}].

\subsection{Low-energy expansion}

Before moving to the solution of the problem at hand, we first review the case of \emph{homogeneous} uniaxial strain~\cite{Pereira2009b}. We take $t_{\bm{r} \bm{r'}}$ as independent of spatial coordinates but possibly dependent on the bond direction $n$,
\begin{align}
t \rightarrow t_n = t e^{-\gamma \Delta u_n} \approx t \left(1 - \gamma \Delta u_n \right).    
\label{eq:hoppings_modulation_linearized}
\end{align} 
Here we expanded the exponential for small displacements
\begin{align}
\Delta u_n = \sum_{i,j} \frac{\delta_n^i \delta_n^j}{a_0^2} \epsilon_{ij},
\end{align}
where the nearest-neighbor vectors $\bm{\delta}_n$ are given by
\begin{equation} 
\bm{\delta_1} = a_0[0,1], ~ \bm{\delta}_2 = \frac{a_0}{2}[-\sqrt{3}, -1], ~ \bm{\delta}_3 = \frac{a_0}{2}[\sqrt{3}, -1].
\end{equation}
The strain tensor  $\epsilon_{ij} = \frac{1}{2} \left[ \partial_j u_i + \partial_i u_j \right] $ is defined through the in-plane displacement field $\bm{u} = (u_x, u_y)$ which we take as a smooth function of the coordinates. In this work we assume that $ \bm{u}$ is only a function of $y$, such that $\epsilon_{xx} = \epsilon_{xy} = \epsilon_{yx} = 0$ and  $\Delta u_1 = \epsilon_{yy}$, $\Delta u_2 = \Delta u_3 = \epsilon_{yy}/4$. 

In momentum space, the Bloch Hamiltonian is given by $h(\bm{k}) = \bm{d}(\bm{k}) \cdot \bm{\sigma} $ where $\bm{\sigma} = (\sigma_x, \sigma_y)$ acts on the sublattice $(A,B)$ degree of freedom and
\begin{align}
    d_x(\bm{k}) &= - t_1 \cos k_y - 2 t_2 \cos \frac{\sqrt{3} k_x }{2} \cos \frac{k_y}{2}, \\
    d_y(\bm{k})&= + t_1 \sin k_y - 2 t_2 \cos \frac{\sqrt{3} k_x }{2} \sin \frac{k_y}{2},
\end{align} 
where $t_n = t(1-\gamma \Delta u_n)$ and we set $a_0 = 1$ from here on. Expanding to lowest-order in momentum around the inequivalent Dirac points $\bm{K}^\pm = ( \pm \frac{4 \pi}{3 \sqrt{3}},0)$, with $\bm{k} = \bm{K}^\pm + \bm{q}$, we obtain
\begin{align}
d^\pm_x(\bm{q}) &= \pm \hbar v_F \left[ (1 - \frac{\gamma}{4}\epsilon_{yy}) q_x \pm \frac{\gamma}{2} \epsilon_{yy}\right]  \label{eq:low-energy_x}, \\
d^\pm_y(\bm{q}) &= \hbar v_F \left[(1 - \frac{3 \gamma}{4} \epsilon_{yy})q_y \pm \frac{q_x q_y}{2} (1 - \frac{\gamma}{4} \epsilon_{yy}) \right],
 \label{eq:low-energy_y}
 \end{align}
where $v_F = 3t/2 \hbar$ is the Fermi velocity and the superscript $\pm$ refers to the two valleys $\bm{K}^\pm$. Note that we expanded to linear order \emph{separately} in both momentum components -- this generates a term proportional to $q_x q_y$ which is usually neglected when considering the low-energy physics of Dirac fermions in graphene. Here it is important because of the broken rotation symmetry of the problem -- when considering non-uniform strain in Sec.~\ref{Sec:LLsol}, only the momentum $q_x$ remains a good quantum number and one must consistently treat all terms linear in $q_x$ to obtain a quantitatively correct result.

As expected, the strain tensor component $\epsilon_{yy}$ couples to $q_x$ as a \emph{pseudo}-gauge field $\mathcal{A}_x = \gamma \epsilon_{yy}/2$ (that is, with a different sign between the two valleys). However, to the same order, $\epsilon_{yy}$ also renormalizes the Fermi velocities along the $q_x$ and $q_y$ directions. In the case of homogeneous strain, this only produces an anisotropic Dirac cone. However, when $\epsilon_{yy}$ is promoted to a function of the coordinates, it has an important effect on the low-energy spectrum~\cite{deJuan2012, Oliva2015}, as described in Sec.~\ref{Sec:LLsol}.

\subsection{Symmetries}

We now briefly comment on the relevant symmetries. Combining the sublattice and valley degrees of freedom, the low-energy limit of the problem is described by the 4-dimensional matrix Hamiltonian
\begin{equation}
    h(\bm{q}) =
    \begin{pmatrix}
    h^+(\bm{q}) & 0 \\
    0 & h^-(\bm{q})
    \end{pmatrix} 
    \label{eq:4matrix},
\end{equation}
where $h^\pm(\bm{q}) = \bm{d}^\pm(\bm{q}) \cdot \bm{\sigma}$ describes the two valleys. We henceforth denote the Pauli matrices acting on the valley pseudospin as $\bm{\tau}$. The fact that the two valleys are decoupled in the low-energy limit allows one to ask meaningful questions about valley transport. Contrary to the case of spin transport (for example), valley is not a microscopic degree of freedom -- strictly speaking, there is no physical symmetry leading to a conserved ``valley charge". Nevertheless, in the effective low-energy description, Eq.~(\ref{eq:4matrix}), one can identify the valley operator $\tau^z \sigma^0$ which is conserved, $[ h(\bm{q}) , \tau^z \sigma^0] =0$, as long as no scattering terms connect the two valleys.

The system respects time-reversal symmetry which acts as $\mathcal{T} h(\bm{q}) \mathcal{T}^{-1} =  \tau^x h^*(\bm{-q}) \tau^x $. In our case this enforces $h^+(q_x) = h^-(-q_x)$ -- that is, the spectra at valleys $\bm{K}^\pm$ are related by a reflection with respect to $q_x=0$. The system also has a chiral (or sublattice) symmetry, $\{ \mathcal{C}, h(\bm{q})$\} = 0 with $\mathcal{C} = \tau^0 \sigma^z$. This implies that the spectrum at \emph{each} valley is symmetric with respect to the charge neutrality point $E=0$. Chiral symmetry is present whenever the terms in $H$ only couple sublattices $A$ and $B$, and will be broken when adding a scalar potential terms in Sec.~\ref{sec:scalar_potential}. Finally, our strain pattern (shown in Fig.~\ref{fig:ribbon}) breaks the inversion symmetry of the lattice, which is a necessary ingredient to generate pseudomagnetic fields.

\section{Exact solution for dispersive pseudo-Landau levels}\label{Sec:LLsol}

\begin{figure*}
\centering
\includegraphics[width=\textwidth]{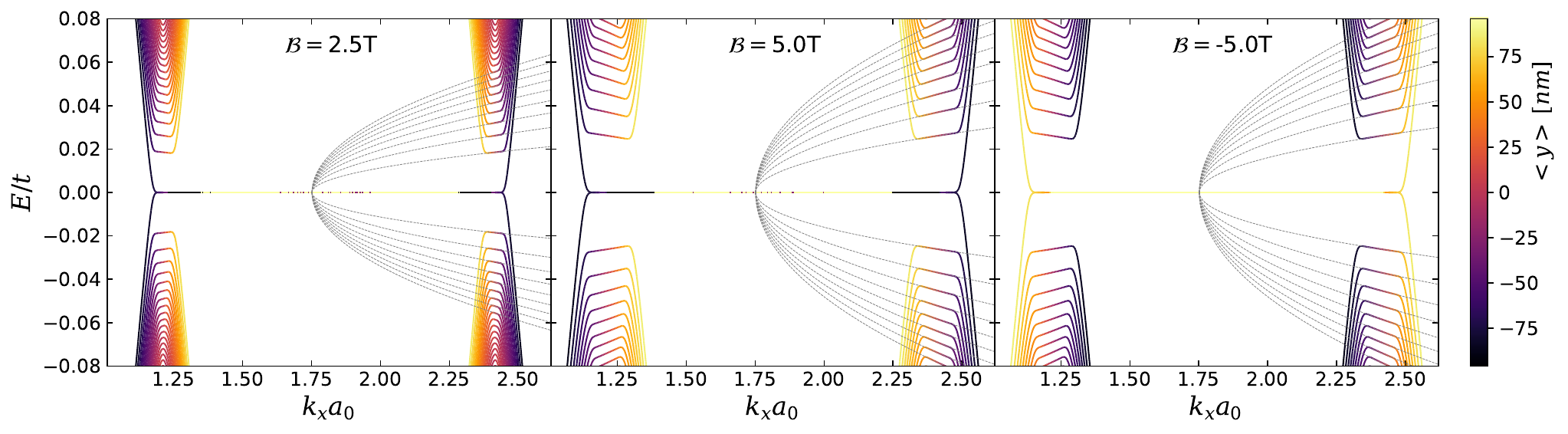}
\caption{Spectra of zigzag graphene nanoribbons (width $W \approx 192$ nm) subject to strain-induced pseudomagnetic fields $\mathcal{B} =2.5$, $5$ and $-5$ T (from left to right). The spectra are obtained from numerical diagonalization of the Hamiltonian [Eqs.~(\ref{eq:tight_binding},\ref{eq:hoppings_modulation_linearized})] and the colorscale represents the expectation value of the $\hat{y}$ position operator for each eigenstate. A set of pseudo-Landau levels with linear dispersion near the Dirac points is visible, as predicted by Eq. (\ref{eq:spectrum_b_plus}) (dotted gray lines, shown here only for valley $\bm{K}^+$). The $n=0$ level remains dispersionless and merges with the usual ``zigzag" zero-energy edge states. Pseudo-Landau levels with higher index $|n|$ can be resolved by increasing the field $\mathcal{B}$. \label{fig:spectrum_B}}
\end{figure*}

Having a low-energy expansion with the correct symmetries in place, we  now promote the strain tensor to a smooth function of the coordinate $y$. This semiclassical substitution is justified if we assume that the displacement field $\bm{u}$ varies slowly on the lattice scale. In order to generate a uniform pseudomagnetic field $\bm{\mathcal{B}} = \mathcal{B} \hat{\bm{z}}$ using only the $\epsilon_{yy}$ component, we take 
\begin{equation}
\epsilon_{yy} = \frac{2 e \mathcal{B} y}{\hbar \gamma} \equiv \frac{2 b y}{\gamma},
\label{eq:strain_definition}
\end{equation}
where we defined the dimensionless parameter $b = e \mathcal{B}/\hbar$. Using Eqs.~(\ref{eq:low-energy_x},\ref{eq:low-energy_y}) the Bloch Hamiltonian in valley $\bm{K}^+$ reads
\begin{align}
 h^+(\bm{q}) = \hbar v_F \left[ \sigma_x \left(q_x +
p by \right) + \sigma_y q_y \left( s - rby\right)  \right]\label{eq:Hamiltonian1}
\end{align}
where we defined $p = 1- q_x/2$, $s = 1 + q_x/2$ and $r = 3/2 + q_x/4$. The corresponding Bloch Hamiltonian $h^-(\bm{q})$ for valley $\bm{K}^-$ is obtained by sending $q_x \rightarrow -q_x$.

By promoting the strain tensor to a function $\epsilon_{yy}(y)$, we have explicitly broken the translational symmetry along $y$. We now perform the canonical substitution $ q_y \rightarrow - i \partial_y$ and, using the remaining translational symmetry along $x$, we look for solutions of $h^+(\bm{q})$ of the form $\Psi(y, q_x) \propto e^{iq_xx} \phi_\alpha(y)$, where $\alpha = A,B$ is the sublattice index. This leads to the following eigenvalue problem,
\begin{align}\label{fff}
\left[ \sigma_x (q_x + pby) - \mathrm{i} \sigma_y (s-rby)\partial_y \right]
\begin{bmatrix}
\phi_A(y) \\
\phi_B(y) 
\end{bmatrix} = \frac{E}{\hbar v_F} \begin{bmatrix}
\phi_A(y) \\
\phi_B(y) 
\end{bmatrix},
\end{align}
with homogeneous Dirichlet boundary condition at infinity. This differential equation differs from the conventional Landau level problem by the presence of terms of the form $y\partial_y$ which complicate the analysis, and also need a proper hermitization. Nevertheless, as discussed in Appendix~\ref{Append:exactsol}, this problem can be solved by transforming Eq.\ \eqref{fff} to a second-order ordinary differential equation for, say, $\phi_B$, 
\begin{align}\label{eq:2nd_ODE_hermi_maintext}
&\left[(pby)^2+\frac{pb}{r}(2q_xr+2ps-br^2)y+\frac{\Delta}{r^2}-\frac{b^2r^2}{4}\right]\phi_B \nonumber \\
&-b^2r^2(2y\phi_B'+y^2\phi_B'')=0 ,
\end{align}
where $\Delta=(q_xr+ps)^2-r^2(\frac{E}{\hbar v_F})^2$.

This turns out to be a \textit{singular} Sturm-Liouville problem~\cite{Boyce2012,Teschl2014} with a regular singularity at $y_\mathrm{sgl}=\frac{s}{rb}$ and an irregular singularity at infinity. It exhibits an unusual finite sequence of eigenvalues differing from that of \textit{regular} Sturm-Liouville problems. The spectrum of Eq.~\eqref{eq:2nd_ODE_hermi_maintext} (and its equivalent result for valley $\bm{K}^-$) is given by
\begin{align}
   \left( E^\pm_n \right)^2 = \hbar^2 v_F^2 \left( |b n|(2\pm3q_x)-[bn(6\pm q_x)/4]^2 \right),
   \label{eq:spectrum_exact}
\end{align}
where 
\begin{equation}
n=0,1,2,\cdots, \left \lfloor\frac{8(2\pm3q_x)}{|b|(6\pm q_x)^2} \right \rfloor
\end{equation}
and $\lfloor a\rfloor$ denotes the greatest integer less than or equal to $a$. The Landau level spectrum is thus \emph{bounded from above}, with more levels attainable at smaller $|b|$. This can be intuitively understood by comparing $|y_\mathrm{sgl}|$ to the typical wavefunction size $\sim l_{\mathcal{B}}$, the magnetic length. When the wavefunction size becomes comparable to $|y_\mathrm{sgl}|$ it is significantly affected by the regular singularity, eventually leading to the breakdown of the pLL spectrum. When $|b|$ decreases, levels with larger $n$ appear in the spectrum because $|y_\mathrm{sgl}| \sim 1/|b|$ grows faster than $l_{\mathcal{B}} = 1/\sqrt{|b|}$. For the strain-induced pseudomagnetic fields considered in this work, one typically has $|b|\sim10^{-4}\ll1$. Thus, $|y_\mathrm{sgl}| \sim 10^4$ is much bigger than both the wavefunction size and the ribbon width $W$, and does not directly influence our analysis.

For small $|b n|$, the quadratic term in Eq.~(\ref{eq:spectrum_exact}) can be safely neglected, leading to
\begin{align}
   E^{\pm}_n = {\rm sgn} (n) \hbar v_F \sqrt{ |bn| (2\pm3q_x) }.
    \label{eq:spectrum_b_plus}
\end{align}
This dispersion relation is peculiar in that the pLLs are not flat, but disperse linearly for small $q_x$ away from the Dirac point, with an opposite slope between the two valleys. When $q_x=0$, the conventional Landau level spectrum is recovered.

We confirm this analytical result by numerically diagonalizing the tight-binding model, Eq.~(\ref{eq:tight_binding}) with hoppings given by Eq.~(\ref{eq:hoppings_modulation_linearized}), and compute the expectation value of the $\hat{y}$ position operator for all eigenstates, as shown in Fig.~\ref{fig:spectrum_B}. Our numerical results confirm the presence of dispersing bulk pLLs described by Eq.~(\ref{eq:spectrum_b_plus}). Landau levels with $n \neq 0$ eventually merge into edge states dispersing upwards (for $n>0$) or downwards (for $n<0$), which are of course not captured by our bulk solution. The bulk pLL$_0$ remains dispersionless and merges with the usual zero-energy zigzag edge states for momenta $q_x$ between the two Dirac cones~\cite{Fujita1996}. The number of bulk pLLs that are resolved depends on the interplay between the magnetic length $l_{\mathcal{B}}$ and the width of the nanoribbon $W$. In our geometry, the pLL wavefunctions have an extent $\sim l_\mathcal{B}$ in the $y$ direction which also increases with $|n|$. Thus, as $|n|$ increases the confining effect from the width $W$ becomes stronger, eventually rendering our bulk solution invalid. Conversely, as the pseudomagnetic field $|b|$ is increased, pLLs with higher $|n|$ can be resolved (see Fig.~\ref{fig:spectrum_B}). 

We note that linearly-dispersing pLLs in uniaxially-strained graphene have been observed numerically in recent works~\cite{He2013,Cazalilla2017}. The linear dispersion was attributed to hybridization with edge modes in Ref.~\onlinecite{He2013}. In contrast, it was argued to be a bulk effect in Ref.~\onlinecite{Cazalilla2017}, using a perturbative treatment of symmetry-allowed terms in the low-energy theory describing the system. Our exact, non-perturbative solution unambiguously identifies the dispersion of the pLLs as a bulk effect. Further, it allows us to obtain a quantitative match with tight-binding simulations by including the (seemingly) higher-order $q_xq_y$ term in the low-energy theory, Eq.~\eqref{eq:low-energy_y}. This term proves to be crucial in obtaining the correct value for the slope of the pLLs near the Dirac points [Eqs. (\ref{eq:spectrum_exact}) and (\ref{eq:spectrum_b_plus})] because it contributes to leading order in the pseudomagnetic field b.

\subsection{Bulk valley currents}

These results have an interesting consequence -- when the chemical potential $\mu$ lies within a pLL, we expect pure valley currents \emph{in equilibrium} in the bulk of the nanoribbon. This is because the two sets of chiral pLLs in valleys $\bm{K}^\pm$ disperse with opposite velocities along $x$. They thus carry only the valley degree of freedom but no electric charge, as shown schematically in Fig.~\ref{fig:ribbon}. We calculate the valley current in equilibrium, as a function of the chemical potential $\mu$, assuming ballistic conduction:
\begin{equation}
    I^v(\mu) = \sum_{s=\pm} \sum_n \int\frac{dq_x}{2\pi} \left[ s f(E_n^s(q_x))  v^s_n(q_x)\right]
    \label{eq:Ivalley_def},
\end{equation}
where $s = \pm$ denotes the two valleys,
\begin{equation}
    v^\pm_n(q_x) \equiv \frac{1}{\hbar}\frac{\partial E^\pm_n}{\partial q_x} = \pm {\rm sgn}(n) \frac{3 v_F}{2} \sqrt{ \frac{|bn|}{2 \pm 3 q_x}}
\label{eq:group_velocity}
\end{equation} 
represents the group velocity of electrons in band $n$, and $f(E^\pm_n) =  1/(e^{(E^\pm_n-\mu)/k_B T}+1 )$ is the Fermi function at temperature $T$. We define $I^v(0) = 0$ as a conventional reference point, noting that the notion of valley is only well-defined close to charge neutrality. At $T=0$ this leads to a contribution $I^v(\mu) = 2 |\mu|/h$ for each pLL.
\begin{figure*}
\centering
\includegraphics[width=\textwidth]{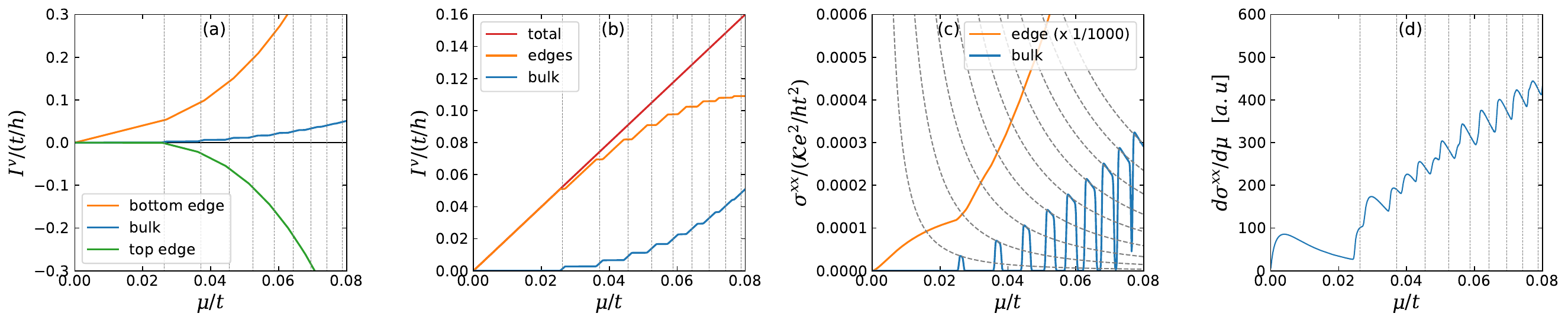}
\caption{Transport properties of strained graphene nanoribbons with $\mathcal{B}=5$ T. (a) Equilibrium valley currents $I^v$  [Eq.~(\ref{eq:Ivalley_def})] as a function of chemical potential $\mu/t$. We separate bulk, top and bottom edge contributions in three equal sections according to the expectation value $\langle \hat{y} \rangle$ of each eigenstate. The vertical lines show the energy of the bulk pLLs at the Dirac point. (b) Same data as in (a), but with the contribution from the two edges combined. The net valley current is non-zero because of the dispersing edge mode connected to the flat, zero-energy $n=0$ pLL. (c) Longitudinal conductivity $\sigma^{xx}$ [Eq.~(\ref{eq:conductivity_def})] as a function of chemical potential, where the edge contribution is suppressed by a factor $1000$. The dashed lines represent the expected bulk conductivity Eq.~(\ref{eq:conductivity_res}) for the first few pLLs with $n=1..9$. (d) Derivative of the total $\sigma^{xx}$ (sum of the edge and bulk contributions in (c)) with respect to $\mu$. We use a small physical temperature $T/t=0.0001$ for all plots.
\label{fig:currents}}
\end{figure*}

We show in Fig.~\ref{fig:currents} the valley currents $I^v(\mu)$ computed using Eqs.~(\ref{eq:Ivalley_def}, \ref{eq:group_velocity}) close to charge neutrality, using the numerical data from our tight-binding calculations. To connect with our bulk solution for the pLLs, we artificially separate the wire in three equal-width regions (along the $y$ direction) which we define as the bottom edge, bulk and top edge, respectively. We compute the contribution to the valley current for each region separately, according to the expectation value $\langle \hat{y} \rangle$ of the corresponding eigenstates  (see Fig.~\ref{fig:currents} a and b). The bulk contribution shows linearly-increasing regions with $I^v(\mu) \sim 2 |\mu| / h$ corresponding to well-formed pLLs, and plateaus when $\mu/t$ lies within a bulk gap, as shown in Fig.~\ref{fig:spectrum_B}. (Note that, contrary to conventional (pseudo)-Landau levels, such bulk gaps only exist here because of the finite width of the ribbon). As expected, the two edges contribute large valley currents of opposite signs, owing to their large group velocity. However, they do not cancel out completely because the $n=0$ edge mode -- connected to the zero-energy, flat zigzag state -- is uncompensated: the $\bm{K}^+$ valley only has a right-moving edge mode and its left-moving partner lives in valley $\bm{K}^-$.

\subsection{Chiral anomaly and negative strain-resistance}

The strain-induced pLL spectrum displayed in Fig.~\ref{fig:spectrum_B} has important consequences for electrical transport which might be the most practical method to probe it experimentally. These include the chiral anomaly and negative strain-resistance which we now discuss.

The structure of the edge modes associated with pLL$_0$, with one right-moving branch belonging to one valley and one left-moving branch to the other, will give rise to the chiral anomaly. The application of a bias voltage \emph{along} the wire will result in pumping electrons from one valley to the other, thus producing a net valley polarization mostly localized at one edge of the ribbon. This is analogous to the chiral anomaly in $(1+1)$ dimensional field theory, where the number of left-moving and right-moving chiral modes is not conserved. Here, because of the oppositely dispersive pLLs at the two valleys, the valley and chiral charge coincides.  Therefore, the valley pumping through the pLL$_0$ is described by the ``valley anomaly" equation
\begin{equation}
\partial_t \rho^v + \partial_x I^v = \frac{e}{h}\epsilon^{\mu\nu}F_{\mu\nu}, 
\end{equation}
where $\rho^v$ and $I^v$ are the valley charge density and valley current, respectively. Note that only the electric field $F_{01}=-F_{10}=E$ exists in $(1+1)$ dimensions. Similarly, in our model the pseudomagnetic field $b$ only determines the slope of the chiral modes, and the one-dimensional transport will be driven solely by an external electric field. Similar to the regular chiral anomaly~\cite{Nielsen1983,Fukushima2008,Zyuzin2012} or its strained-induced counterpart~\cite{Pikulin2016,Grushin2016} in $(3+1)$ dimensional Weyl/Dirac semimetals, we may expect a large contribution to the conductivity from the valley anomaly. This is because the imbalance between the valley charge densities can only be relaxed through intervalley scattering whose rate tends to be suppressed due to the large momentum space separation between the valleys.  

A unique manifestation of the pLL spectrum is the negative longitudinal ``strain-resistance", which results from the group velocity of the $n$th Landau level being both $b$ and $n$ dependent as indicated in Eq.~(\ref{eq:group_velocity}).  The longitudinal DC conductivity for Landau level $n$, in the semiclassical Boltzmann formalism, is given by
\begin{equation}
    \sigma_n^{xx} = e^2 \int \frac{dq_x}{2\pi} \tau(E_n(q_x)) v^\pm_n(q_x)^2 \left. \left( -\frac{\partial f(E - \mu)}{\partial E} \right)  \right|_{E_n(q_x)}
    \label{eq:conductivity_def}
\end{equation}
where $\tau(E_n (q_x))$ is the relaxation time and $\mu$ is the Fermi energy. Using Eq.~(\ref{eq:group_velocity}) for the group velocity of electrons in the bulk pLLs and changing the integration variable to band energy $E_n$, we obtain (at zero temperature)
\begin{align}
    \frac{\sigma_n^{xx}}{\left(e^2/h\right)} 
    &= \frac{3 \hbar v_F^2 |bn| \tau(\mu)}{2 \mu}.
\end{align}
In real systems $\tau(\mu)$ will be a phenomenological parameter describing various contributions to electronic scattering. In the following we assume for simplicity that the dominant source of scattering is a Drude contribution which can be treated using the Born approximation. The relaxation time then reads $\tau(\mu) = \hbar / 2 \pi D(\mu) n_{i} C$ where $n_{i}$ is the concentration of impurities, $C$ is a constant depending on the details of the scattering and $D(\mu)^{-1} = 2 \pi \hbar |v_n^\pm(\mu)|$ is the density of electronic states at the Fermi level $\mu$. We obtain
\begin{equation}
\tau(\mu) = \frac{\hbar^2 |v_n^\pm (\mu)|}{ n_{i} C} = \frac{3 \hbar^3 v_F^2}{2 n_{i} C} \frac{ |bn|}{\mu}
\end{equation}
and thus
\begin{align}
    \frac{\sigma_n^{xx}}{\left(e^2/h\right)} 
    = \frac{ \mathcal{K} |bn|^2 }{\mu^2}.
    \label{eq:conductivity_res}
\end{align}
where we defined the constant $\mathcal{K} = 9 (\hbar v_F)^4/ 4 n_{i} C$. We see that the longitudinal resistivity $\rho^{xx} = 1/\sigma^{xx}$ of the pLLs decreases as $1/|b|^2$, giving rise to a characteristic {\em negative strain resistance}. Note also the peculiar $\mu$ dependence reflecting the fact that the pLLs are not linear far from the Dirac points.

In systems with a finite width, the pLLs are not completely formed -- they only exist in a region of momentum space near the Dirac points. Consequently, the bulk expression for $\sigma_n^{xx}$ only applies for $\mu$ inside a well-formed bulk pLL, as shown in Fig.~\ref{fig:currents}c, which compares the tight-binding simulation with the analytical prediction, Eq.~(\ref{eq:conductivity_res}). The edge states (not captured by this argument) provide the dominant contribution to $\sigma^{xx}$, given their large group velocity, and might therefore mask the contribution from the bulk pLLs in experiments. Nevertheless, clear signatures of the pLLs can be seen in $d\sigma^{xx}/d\mu$, which shows oscillations with a series of minima between two neighboring peaks (see Fig.~\ref{fig:currents}d). These peaks correspond to the abrupt change in group velocity occurring when a bulk pLL merges into an edge state.

\section{Scalar potential}
\label{sec:scalar_potential}

\begin{figure*}
\centering
\includegraphics[height=4.8cm]{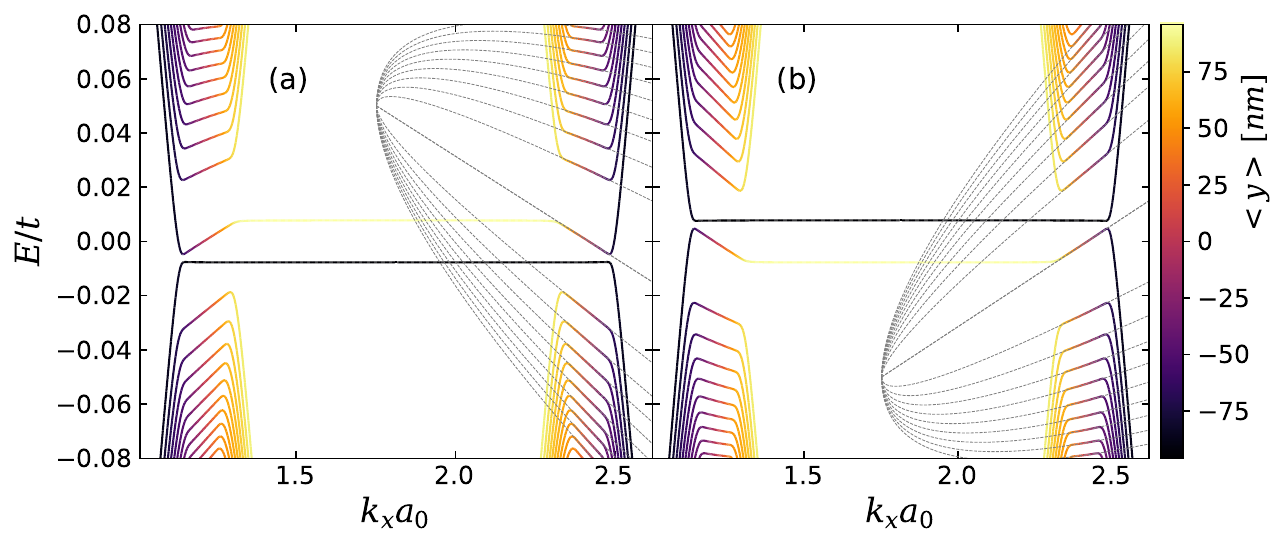}
\includegraphics[height=4.8cm]{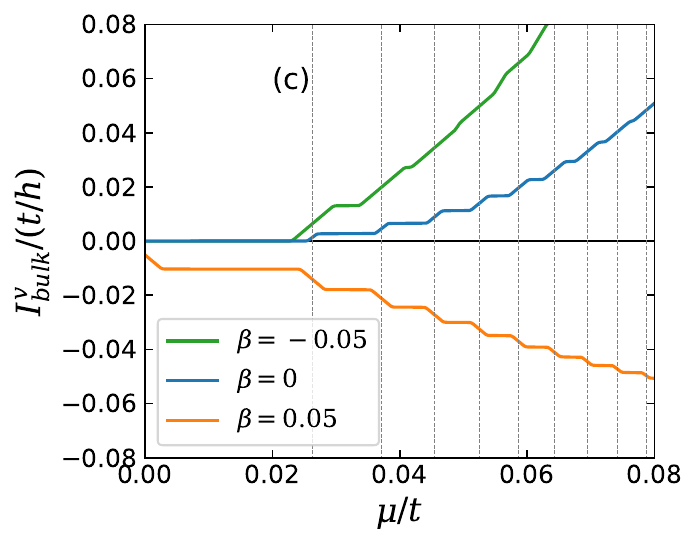}
\caption{Spectra of zigzag graphene nanoribbons (width $W \approx 192$ nm) subject to pseudomagnetic field $\mathcal{B}=5$ T and electric fields with $\beta = E/v_F \mathcal{B} = 0.05$ (a) and  $-0.05$ (b). The spectra are obtained from the exact-diagonalization of the Hamiltonian [Eqs.~(\ref{eq:tight_binding},\ref{eq:hoppings_modulation_linearized}) with the addition of the scalar potential term, Eq.~(\ref{eq:scalar_potential_lattice})], and the colorscale represents the expectation value of the $\hat{y}$ position operator for each eigenstate. The dispersion of the pLLs follows the prediction based on the solution of the Dirac equation in the boosted frame, Eq.~(\ref{eq:spectrum_b_e}) (dotted gray lines, shown here for valley $\bm{K}^+$ only). The $n=0$ pLL now disperses with an opposite slope in the two valleys. (c) Bulk contribution to the equilibrium valley current for different $\beta$. \label{fig:spectrum_be}}
\end{figure*}

In general, a non-uniform deformation of the lattice will not only generate a pseudomagnetic field, but also a \emph{scalar} potential energy $U$. This was so far neglected by considering only changes in hopping parameters. In general, such a scalar potential term is proportional to the trace of the strain tensor~\cite{Suzuura2002,Manes2007},
\begin{align}
U = \lambda \left( \epsilon_{xx} + \epsilon_{yy} \right),
\end{align}
where $\lambda$ is a coupling constant which depends on microscopic details and is estimated to be $\lambda \sim 4 ~ \text{eV}$ in monolayer graphene~\cite{Castro2017}. For the simplest tri-axially symmetric strain pattern originally considered in Ref.~[\onlinecite{Guinea2009}], $\epsilon_{xx} = -\epsilon_{yy}$ and $U$ vanishes by symmetry. However, for the uniaxial strain considered here, this is not the case and $U = 2 \lambda b y/\gamma$. In our tight-binding model, this term simply reads
\begin{equation}
    H_{\text{scalar}} = \sum_{\bm{r}} U \left( a^\dagger_{\bm{r}} a_{\bm{r}} + b^\dagger_{\bm{r}} b_{\bm{r}} \right),
\label{eq:scalar_potential_lattice}
\end{equation}
which breaks chiral symmetry but respects time-reversal symmetry. It has the same form as the coupling to an externally applied electric field $\bm{E}  = (E \bm{\hat{y}} = 2 \lambda b/e \gamma) \bm{\hat{y}}$. Thus, even though the parameter $\lambda$ cannot be directly controlled, it can be canceled (or enhanced) by applying a bias voltage across the nanoribbon. The presence of this scalar potential term was argued~\cite{Castro2017} to explain some features of the pLL spectrum observed in Ref.~[\onlinecite{Levy2010}].  

\subsection{Solution for pseudo-Landau levels}

In the presence of the scalar potential term, the low-energy theory for valley $\bm{K}^+$ [Eq.~(\ref{eq:Hamiltonian1})] becomes
\begin{align}
h^+(\bm{q}) = \hbar v_F \left[ \sigma_x \left(q_x + p \mathcal{A}_x \right) + \sigma_y q_y \left( s - rby\right) + \sigma_{0} A_0 \right].
\label{eq:h_electricfield}
\end{align}
where we defined the electromagnetic potential $A_\mu = (A_0, \mathcal{A}_x, 0)$ with $A_0 = e E y/ \hbar v_F$ and $\mathcal{A}_x = by = e \mathcal{B}y/\hbar$. This Hamiltonian is reminiscent of the seminal problem of a 2+1-dimensional massless Dirac fermion in perpendicular electric and magnetic fields, which can be solved in an elegant manner by boosting to a frame where the electric field vanishes~\cite{Lukose2007, Peres2007}. In our case, a complication arises: the Dirac equation obtained from Eq.~(\ref{eq:h_electricfield}) is not Lorentz invariant because $p$, $r$, and $s$ are not Lorentz scalars but functions of the spatial derivative $\partial_x$. However, when considering small pseudomagnetic fields and momenta near the Dirac points (such that $q_x \ll a_0^{-1}$) these Lorentz-invariance breaking terms are small. One is thus tempted to treat $p$, $r$ and $s$ as Lorentz scalars and derive an approximate expression for the Landau level spectrum in the presence of both $\bm{E}$ and $\bm{\mathcal{B}}$. We then confirm that this method yields a quantitatively correct spectrum for small fields, near the Dirac points, by comparing directly with tight-binding results.

With this is mind, we now perform a Lorentz boost along the $x$ direction to a new frame $\tilde{S}$, following Ref.~[\onlinecite{Lukose2007}]. The coordinates of the new frame are given by $\tilde{x^\nu} = \Lambda_\mu^\nu x^\mu$, where $x^\mu = (v_F t, x, y)$ and
\begin{align}
\Lambda = \begin{pmatrix}
\cosh \theta & \sinh \theta & 0 \\
\sinh \theta & \cosh \theta & 0 \\
0 & 0 & 1
\end{pmatrix}.
\end{align}
The relative velocity $\beta$ between the old and the new frame is determined through the usual relationship $\tanh \theta = \beta$. The electromagnetic potential transforms with the inverse transformation as $\tilde{A}_\nu = \Lambda^\mu_\nu A_\mu$ which yields
\begin{align}
\tilde{A}_0 &= \cosh \theta \left(\frac{E}{v_F \mathcal{B}} - \beta \right) b y, 
\nonumber \\
\tilde{A}_x &= \cosh \theta \left(1 - \frac{\beta E}{v_F \mathcal{B}} \right) b y .
\end{align}
Choosing the new frame velocity as $\beta = E/v_F \mathcal{B}$, the electric field vanishes and we are left with a problem of the same form as in Eq.~(\ref{eq:Hamiltonian1}), albeit with a renormalized pseudomagnetic field $\tilde{b} = b \sqrt{1-\beta^2}$. Invoking our previous result [Eq.~(\ref{eq:spectrum_b_plus})], the solution for the pLLs in the new frame (to lowest-order in the field $b$) reads
\begin{align}
   \tilde{E}^+_n = {\rm sgn}(n) \hbar v_F (1-\beta^2)^{1/4} \sqrt{|b n|(2 + 3\tilde{q}_x) }.
\end{align}
Transforming back to the original frame (see details in Appendix \ref{Append:frame_transformation}) we obtain, to lowest order in $|b|$, $\beta$ and $q_x$,
\begin{align}
   \frac{E^+_n}{\hbar v_F} =& -\beta q_x + \frac{3}{2} \beta |b n| \nonumber \\
   &+{\rm sgn}(n) (1-\beta^2)^{3/4} \sqrt{|b n|(2 + 3 q_x)}.
    \label{eq:spectrum_b_e}
\end{align}
In the $\bm{K}^-$ valley, the spectrum is obtained by the replacement $q_x \rightarrow - q_x$ to respect time-reversal symmetry. 

Numerical simulations on the strained lattice model incorporating the scalar potential term Eq.~(\ref{eq:scalar_potential_lattice}) reproduce the analytical result Eq.~(\ref{eq:spectrum_b_e}) for the bulk modes, as shown in Fig.~\ref{fig:spectrum_be}. The most striking effect of taking the electric field into account is that LL$_0$ also acquires a linear dispersion with an opposite slope $\mp \beta \hbar v_F$ in the two valleys. Therefore, LL$_0$ contributes to bulk valley currents under the combination of $\bm{E}$ and $\bm{\mathcal{B}}$ fields, whereas LL$_n$ with $n \neq 0$ only require the presence of $\bm{\mathcal{B}}$.

\subsection{Consequences for valley currents and tunable flat bands}

In the presence of the scalar potential term, the group velocity of bulk pLLs changes to
\begin{equation}
    v^\pm_n(q_x) = \mp \beta v_F \pm {\rm sgn}(n) \frac{3 v_F}{2} (1 - \beta^2)^{3/4}\sqrt{ \frac{|bn|}{2 \pm 3 q_x}}
\label{eq:group_velocity_E}
\end{equation} 
which affects the valley currents and can also change their sign, as shown in Fig.~\ref{fig:spectrum_be}. This mechanism provides a way to electrically control the magnitude as well as switch the polarity of the bulk valley currents. Similarly, the longitudinal conductivity $\sigma^{xx}$ will be affected by the slope renormalization (not shown here).

Another interesting feature of this system is that by tuning the strength of the electric and magnetic contributions, the linear in $q_x$ part of the dispersion can be canceled out for any given Landau index $n$, thus generating a nearly flat band. This could be accomplished by tuning the electric field across the nanoribbon with bias voltages such that
\begin{equation}
   \frac{ \beta }{\left(1-\beta^2\right)^{3/4}} = {\rm sgn}(n)  \frac{3}{2} \sqrt{\frac{|bn|}{2}}.
\end{equation}
This mechanism could thus provide a way to tune correlation effects in graphene nanoribbons.

\subsection{Quantitative estimates}

Here we provide estimates of quantities relevant for the experimental exploration of the valley physics discussed in this work. First, it is crucial that pLLs are formed in the system. This necessitates a magnetic length that is much smaller than the width $W$ of the ribbon,
\begin{equation} 
l_{\mathcal{B}} = \sqrt{\frac{\hbar}{e \mathcal{B}}} \ll W.
\end{equation}
For $\mathcal{B} = 2.5$T ($\mathcal{B} = 5$T) shown in Fig.~\ref{fig:spectrum_B}, we get $l_{\mathcal{B}} \sim 16$ nm ($l_{\mathcal{B}} \sim 11.5$ nm) as a lower bound for $W$. 
For a linearly-increasing strain with the unstrained (equilibrium) point located in the middle of the nanoribbon, the maximal relative displacement $\Delta u_{\text{max}}$ will be experienced at the edges ($y = \pm W/2$) and given by
\begin{equation}
    \Delta u_{\text{max}} \sim \epsilon_{yy}(y=\frac{W}{2}) 
    = \frac{e a_0 \mathcal{B}}{\gamma \hbar} W ,
\end{equation}
where we restored $a_0$. For $\mathcal{B} = 2.5-5.0$T and $W \sim 192$ nm as shown in Fig.~\ref{fig:spectrum_B}, we get $\Delta u_\text{max} \sim 3-6$ \%. These values are not unreasonable, as Ref.~[\onlinecite{Schonenberger2019}] reports that in-situ uniaxial strain gradients of $\sim1 \%$ can already be created. They are also well below the $\sim 20 \%$ threshold which monolayer graphene can withstand without breaking~\cite{Lee2008}. Further, the characteristic energy gaps induced for fields $B = 2.5 - 5.0$ T are given (at the Dirac point) by $E_{gap} \sim 55-78$ meV and are thus within the experimental resolution of ARPES or STM techniques.

Finally, our calculation relies on $\beta < 1$ where $\beta=1$ correspond to the collapse of the pLL spectrum. For $\mathcal{B}=5$ T and $v_F \approx 9 \times 10^5 \text{m/s}$ in graphene, this leads to the condition $E < 4.5 \times 10^3$ V/mm. In comparison, we get $E = 2 \lambda a_0 \mathcal{B}/ \hbar \gamma \approx 2.5 \times 10^3$ V/mm, indicating that this system is close to the pLL collapse (as also discussed in Ref.~[\onlinecite{Castro2017}]). However, in our setup $E$ can be controlled by applying bias voltages across the ribbon, thus allowing to access the small $\beta$ regime discussed in this work.

\section{Discussion and Outlook}
\label{Sec:discussion}

In this paper we presented a simple setup which generates spatially separated valley currents and a valley analog to the chiral anomaly in uniaxially-strained graphene nanoribbons. These features are a direct consequence of dispersive bulk pLLs near the Dirac points. We showed how this anomalous dispersion arises through an inhomogeneous Fermi velocity, which naturally appears in the low-energy theory describing our system, and which lends itself to an exact analytical solution. The effects of an applied electric field were also considered by generalizing the solution of Ref.~[\onlinecite{Lukose2007}] to our setup, showing how the valley currents can be controlled (and even reversed) by bias voltages applied across the ribbon.

We now conclude by providing further remarks relevant to potential experimental realizations of the physics discussed in this work. One major challenge will be to engineer such a non-uniform, uniaxial strain pattern in a real graphene nanoribbon. This differs from the proposal of Ref.~[\onlinecite{Zhu2015}], which employs a uniaxial \emph{stretch} to generate a uniform pseudomagnetic field which, however, is not equivalent to a uniaxial strain due to the particular geometry used. One solution could be to bend a graphene nanoribbon (or a flexible substrate on which the graphene nanoribbon would be deposited) in a spiral-like shape, where the radius of curvature $r(\theta)$ would depend linearly on the angle $\theta$. Perhaps more practically, a recent work~\cite{Schonenberger2019} reports the creation of linearly-increasing uniaxial strain patterns similar to those considered in this work. Using in-situ strain-tuning of graphene encapsulated in hBN, the authors report a maximal strain around $\sim 1 \%$ which is not too far from the $3-6 \%$ used in this work.

Another challenge will be the \emph{detection} of such valley currents. This could potentially be accomplished by attaching a ``valley filter"~\cite{Rycerz2007, Pereira2009, Zhai2010, Fujita2010, Gunlycke2011, Settnes2016b, Stegmann2018} or a ``valley beam splitter"~\cite{Garcia-Pomar2008,  Settnes2016b, Stegmann2018, Prabhakar2019} at one end of the nanoribbon. However, care must be taken in separating the bulk contribution from the edge contributions which tend to dominate transport properties, owing to their large group velocities. Further, careful matching between the filter or beam splitter characteristic energy window and the energy of the pLL under study must be acheived. Using instead a superlattice of valley filters might offer better energy tunability \cite{Torres2019}, but still remains an experimental challenge.

The simplest experimental detection of the anomalous pLL structure predicted in this work can be achieved through an ordinary charge transport measurement. As implied by Eq.\ \eqref{eq:conductivity_res} and Fig.~\ref{fig:currents}, the bulk longitudinal electrical conductivity $\sigma^{xx}$ of a strained nanoribbon exhibits several remarkable features. These include a negative strain-resistance (that is, decreasing resistance with increasing strain) as well as a characteristic non-monotonic dependence on the chemical potential which can be tuned by adjusting the gate voltage. Given that the edge contribution to $\sigma^{xx}$ typically dominates over the bulk contribution, such effects might easier to observe in the derivative of the total conductivity, $d\sigma^{xx}/d\mu$, as shown in Fig.~\ref{fig:currents}d. The unique signatures of pLLs could also be observable using standard spectroscopic techniques including the scanning tunneling spectroscopy and angle-resolved photoemission, which have been successfully employed to probe strain-induced gauge fields in graphene~\cite{Levy2010,Yeh2011,Lu2012,Li2015,Liu2018,Nigge2019}.  

Finally, we remark that our analysis does not crucially rely on a nanoribbon geometry -- it could also be realized as a uniaxally-strained region embedded in a larger (unstrained) graphene sheet. In that case the ``bulk" valley currents would occur in the center of the strained region, and the counter-propagating ``edge" valley currents would be mostly localized at the interfaces between the strained and unstrained regions. This might provide an alternative platform to test the valley physics presented in this work.

\section{Acknowledgments}

The authors are indebted to Anton Burkov, Anfanny Chen, Hiroaki Ishizuka and Tianyu Liu for helpful discussions. This research was supported in part by NSERC, CIfAR, the Heising-Simons Foundation, the Simons Foundation, and National Science Foundation Grant No. NSF PHY-1748958.

\bibliography{biblio}

\appendix

\section{Exact solution of the dispersive pseudo-Landau levels}\label{Append:exactsol}
\subsection{The differential equations and mathematical properties}
For the low-energy physics around Dirac points of pLLs due to spatial modulations in the $y$ direction (Landau gauge), one can write down a most general system of linear ordinary differential equations (ODE) of the $2\times2$ Hamiltonian, $\vec{d}\cdot\vec{\sigma}\begin{bmatrix}  \phi_A(y) \\ \phi_B(y)  \end{bmatrix} = \epsilon  \begin{bmatrix}  \phi_A(y) \\ \phi_B(y)  \end{bmatrix}$ with 
\begin{align}
d_x=-\mathrm{i}(u-vby)\partial_y+pby+q \\
d_y=-\mathrm{i}(s-rby)\partial_y+wby+t
\end{align}
where $\epsilon=E/(\hbar v_F)$ and $p,q,r,s,u,v,w,t$ are certain linear polynomials of $q_x$ dependent on the specific model. It lacks a straightforward analytic solution to the best of our knowledge, unless $u=v=0$ or $r=s=0$ or $s-rby \propto u-vby$, which can all be solved in a way similar to the following. The analytic tractability thus lies in the absence of differentiation in one of $d_x,d_y$ or the presence of a same type of differentiation in both, which, interestingly, always introduces a regular singularity in $(-\infty,\infty)$ as shown below. 

Without loss of generality, we focus on the case when $u=v=0$ to account for all the models considered here
\begin{equation}\label{eq:1st_ODE}
    \vec{d}\cdot\vec{\sigma}=( q+pby) \sigma_x - \mathrm{i}(s-rby)\partial_y \,\sigma_y.
\end{equation}
In the main text, $q$ is replaced directly by the momentum $q_x$. As aforementioned, adding nonzero $w,t$ is still analytically solvable in a similar manner.
For completeness, we mention that two independent artificial modulations could possibly introduce two different rates of modulation, $\bar{b}=\nu b$ and $b$ corresponding respectively to the two $b$'s in Eq.~\eqref{eq:1st_ODE}, although the overall effect is a single pseudomagnetic field dependent on both. Obviously, $\nu$ can be absorbed into $p$. This, in some cases, contrary to the one in the main text, can in its own right generate dispersionless flat pLLs even in zigzag graphene ribbon.

For Eq.~\eqref{eq:1st_ODE}, let's first shift $y\mapsto y+\frac{s}{rb}$ to get 
\begin{equation}
[ q+pb(y+\frac{s}{rb})]\sigma_x + \mathrm{i}\,rby\partial_y \,\sigma_y.    
\end{equation}
Under the usual homogeneous Dirichlet boundary condition
, we note that $-\mathrm{i}y\partial_y$ is not Hermitian. Therefore, one had better first perform a proper \textit{hermitization}, which is not unique in general. A convenient choice that preserves the eigenequation structure is to use instead the symmetrically Hermitized operator $-\mathrm{i}\frac{y\partial_y+\partial_yy}{2}=-\mathrm{i}(y\partial_y+\frac{1}{2})$
. Further eliminating $\phi_A$, we arrive at a 2nd-order ODE \begin{align}\label{eq:2nd_ODE_hermi}
&\left[(pby)^2+\frac{pb}{r}(2qr+2ps-br^2)y+\frac{\Delta}{r^2}-\frac{b^2r^2}{4}\right]\phi_B \nonumber \\
&-b^2r^2(2y\phi_B'+y^2\phi_B'')=0,
\end{align}
where we define $\Delta=(qr+ps)^2-r^2\epsilon^2$. This Hermitization only introduces the last term in the bracket in front of $\phi_B$ and does not alter the overall form of the equation and hence the eigenspectrum. 

Equation \eqref{eq:2nd_ODE_hermi} can be cast in the form of a \textit{singular Sturm-Liouville (SL) problem} 
\begin{equation}
(P\phi_B')'-Q\phi_B=-\lambda W\phi_B    
\end{equation}
with $P(y)=y^2,W(y)=1,Q(y)=[(pby)^2+\frac{pb}{r}(2qr+2ps-br^2)y+\frac{(qr+ps)^2}{r^2}-\frac{b^2r^2}{4}]/(b^2r^2),\lambda=\frac{\epsilon^2}{b^2r^2}$. It is \textit{singular} because the interval is infinite and $P(0)=0$. Therefore, it is not guaranteed to have square integrable (i.e., inside the physical Hilbert space) eigenfunctions on $(-\infty,0]$ or $[0,\infty)$ or $(-\infty,\infty)$ for real eigenvalues, depending on Weyl's spectral dichotomy, limit circle/limit point classification, of the boundary or singular points $0,\pm\infty$. Neither is an infinite sequence of discrete eigenvalues guaranteed~\cite{Boyce2012,Teschl2014}. A finite sequence of discrete eigenvalues is actually what we will see in Eq.~\eqref{eq:eigenvalues}. Knowing that each eigen-subspace is at most one-dimensional for separated boundary conditions and a well-posed SL problem should have a complete eigenbasis~\cite{Teschl2014}, the existence of a complementary continuous essential eigenspectrum above the maximal discrete eigenvalue is naturally expected.

\subsection{Analysis of the equation}
To facilitate the analytic solution, we first perform the asymptotic analysis. Now the original regular singularity $y_\mathrm{sgl}=\frac{s}{rb}$ is moved to $y_\mathrm{sgl}=0$ while the irregular singularity is still at $\infty$. In the vicinity of $y_\mathrm{sgl}$, we can neglect any term dependent on $y$ in the polynomial factor of $\phi_B$ and Eq.~\eqref{eq:2nd_ODE_hermi} becomes
\begin{equation}
\left(\frac{\Delta}{r^2}-\frac{b^2r^2}{4}\right)\phi_B-b^2r^2(2y\phi_B'+y^2\phi_B'')=0.    
\end{equation}
Physically for small momentum $k$ relative to the Dirac point and lower Landau level energies $\epsilon$, we hereby safely assume $\Delta\geq0$ and will justify later. This Cauchy-Euler equation has two independent solutions $y^{-\frac{1}{2}\pm\frac{\sqrt{\Delta}}{|b|r^2}}$. To make the solution not divergent at $y_\mathrm{sgl}$, only $y^{-\frac{1}{2}+\frac{\sqrt{\Delta}}{|b|r^2}}$ is physically acceptable. Towards the infinity, Eq.~\eqref{eq:2nd_ODE_hermi} is asymptotically expressed as
\begin{equation}
(pby)^2\phi_B-(bry)^2\phi_B''=0,    
\end{equation}
which has two independent solutions $\mathrm{e}^{\pm \frac{p}{r}y}$. Note that they diverge at $\pm\infty$, respectively. Therefore, they cannot help build any physical solution on the whole $y$-axis, which is a peculiarity of the present singular SL problem and will be made clear later. 

We are now ready to make the substitution $\phi_B(y)=\mathrm{e}^{-\frac{z}{2}}y^{-\frac{1}{2}+\frac{\sqrt{\Delta}}{|b|r^2}}u(y)$ with a change of variable $z=-\mathrm{sgn}(b)\frac{2p}{r}y$ for brevity. It may look \textit{a priori} for the $\mathrm{sgn}(b)$-dependence and surely can be motivated by a numerical solution. (Although it later helps make the solution mathematically and physically clear, we can otherwise stick to the same substitution without $\mathrm{sgn}(b)$ and get some seemingly distinct solution, which can be shown equivalent by certain transformation properties.) The new equation turns out to be 
\begin{align}
&-|b|\mathrm{e}^{-\frac{z}{2}}y^{\frac{1}{2}+\frac{\sqrt{\Delta}}{|b|r^2}} \times \nonumber \\ &\left\{\frac{p}{r}\left[2\mathrm{sgn}(b)(\sqrt{\Delta}-qr-ps)+r^2(b+|b|)\right]u(y) \right.  \nonumber \\ 
&\left.+ \left(2prby+2\sqrt{\Delta}+r^2|b|\right)u'(y)  + r^2|b|yu''(y) \right\}=0.
\end{align}
Away from the singularity, we further transform it to a \textit{confluent hypergeometric equation} of $u(z)$ 
\begin{equation}
zu''(z)+(\gamma-z)u'(z)-\alpha u(z)=0,    
\end{equation}
in which $\alpha=\frac{(b+|b|)r^2+2(\sqrt{\Delta}-qr-ps)}{2|b|r^2}$ and $\gamma=1+\frac{2\sqrt{\Delta}}{|b|r^2}$. 
Formally, it has two linearly independent solutions, Kummer's function $M(\alpha,\gamma,z)$ and Tricomi's function $U(\alpha,\gamma,z)$~\cite{AbramowitzStegun}. Solution $M$ exists when $\gamma$ is not a non-positive integer, which is manifestly satisfied. Solution $U$ in general exists as a linear combination $\frac{\Gamma(1-\gamma)}{\Gamma(\alpha+1-\gamma)}M(\alpha,\gamma,z)+\frac{\Gamma(\gamma-1)}{\Gamma(\alpha)}z^{1-\gamma}M(\alpha+1-\gamma,2-\gamma,z)$ or only the second part if $\gamma$ is a non-positive integer. Exhausting all the special cases, $U(\alpha,\gamma,z)$ either always contains a term proportional to $z^{1-\gamma}$ or is reduced to $M(\alpha,\gamma,z)$ when $\alpha$ is a non-positive integer and $\gamma$ is not. 

\subsection{Eigenenergy and wavefunction} 
Thus, if solution $U$ were present, $\phi_B$ would have a part $\propto y^{-\frac{1}{2}-\frac{\sqrt{\Delta}}{|b|r^2}}$ that diverges at $y_\mathrm{sgl}$ and hence we can hereafter work only with solution $M$. For $M(\alpha,\gamma,z)$ to not diverge at infinity, it is cut off to the \textit{generalized Laguerre polynomial} $L_{-\alpha}^{\gamma-1}(z)$ when $\alpha$ is a non-positive integer. This solves the eigenvalues
\begin{equation}\label{eq:eigenvalues}
\epsilon^2=|b|n\left(2qr+2ps-|b|nr^2\right),    
\end{equation}
in which $n=0,1,2,\cdots,\lfloor\frac{qr+ps}{|b|r^2}\rfloor$. This already takes into account that the overall power of $y$ in $\phi_B(y)$ should be non-negative for the convergence at $y_\mathrm{sgl}$. And in fact, we have $n\geq1$ ($n\geq0$) when $b>0$ ($b<0$), which is intentionally corrected since this apparent asymmetry between two opposite directions of magnetic field is purely an artifact of converting the 1st-order matrix ODE to a single 2nd-order ODE. Note that the eigenvalues clearly show a higher order effect of the magnetic field $b$ and have an upper bound as seen from the fact that the quantization condition automatically excludes $\Delta<0$. This eigenenergy upper bound, together with the wavefunction expanse (see discussion below), $(-\infty,\frac{s}{rb})$ and $(\frac{s}{rb},+\infty)$ for positive and negative $b$ respectively, diverges with vanishing $b$ as understood from the limiting case recovering the leading-order conventional Landau level. Also note that in Eq.~\eqref{eq:2nd_ODE_hermi}, only one term $\propto b^2y$ in the factor of $\phi_B$ will flip its sign by the mapping $b\mapsto-b,y\mapsto-y$. Therefore, for the opposite direction of magnetic field, the solution of identical energy is not obtained as a simple coordinate reflection with respect to the singularity point $y_\mathrm{sgl}$.

It is also worthwhile to further comment on the wavefunction. Firstly, similar to conventional Landau levels, for $n$th eigenenegy, $\phi_B(y)$ possesses exactly $n-1$ zeros as seen from the Laguerre polynomial contained. Note that this is in general not granted since we have a singular SL problem. Secondly, the exponential factor in $\phi_B(y)$ will diverge at $\mathrm{sgn}(b)\infty$. It looks as if the analytic solution doesn't allow a physical solution vanishing towards both directions of the infinity simultaneously. This is resolved by noticing that homogeneous Dirichlet boundary condition is automatically satisfied at $-\mathrm{sgn}(b)\infty$ and $y_\mathrm{sgl}$. Therefore, the true wavefunction is 
\begin{equation}
\phi_B(y)\theta[\mathrm{sgn}(b)(y_\mathrm{sgl}-y)].    
\end{equation}
This is continuous but not its derivative, which is valid for a wavefunction and is a typical consequence of the regular singularity $y_\mathrm{sgl}$. 
The singularity suggests that some otherwise small term cannot be ignored in the vicinity of the singularity, and adding it eliminates the singularity. Physically, the lattice regularization surely plays such a role. As a final check, we mention that all the features discussed are confirmed by numerical solutions based on a second order finite element method. 

\section{Spectrum of pseudo-Landau levels in the presence of electric and pseudomagnetic fields} \label{Append:frame_transformation}

The Landau level spectrum in the new frame $\tilde{S}$, in the $\bm{K}^+$ valley, reads
\begin{align}
   \tilde{E}_n = {\rm sgn}(n) \hbar v_F (1-\beta^2)^{1/4} \sqrt{|b n|(2 + 3 \tilde{q}_x)},
\end{align}
To obtain the spectrum in the original frame $S$, we have to consider the transformation of the energy-momentum four-vector under the Lorentz transformation $\Lambda$,
\begin{align}
    \tilde{E}_n &= E_n \cosh \theta + \hbar v_F q_x \sinh \theta \label{eq:Lorentz_e} ,\\
    \hbar v_F \tilde{q}_x  &= E_n \sinh \theta + \hbar v_F q_x \cosh \theta ,
    \label{eq:Lorentz_q}
\end{align}
where $\tanh \theta = \beta = E/v_F \mathcal{B}$. Applying first Eq.~(\ref{eq:Lorentz_e}) we obtain
\begin{align}
   \frac{E_n}{\hbar v_F} = -\beta q_x + {\rm sgn}(n) (1-\beta^2)^{3/4} \sqrt{|b n|(2 + 3 \tilde{q}_x)},
\end{align}
which recovers the result of Ref.~[\onlinecite{Lukose2007}] when $\tilde{q}_x = 0$. Transforming $\tilde{q}_x$ to the original frame, using Eq.~(\ref{eq:Lorentz_q}), yields
\begin{align}
   \frac{E_n}{\hbar v_F}
   =& - \beta q_x \\ 
   &+  {\rm sgn}(n) (1-\beta^2)^{3/4}\sqrt{|b n|\left(2 +  3\frac{q_x + \beta E_n/\hbar v_F}{\sqrt{1-\beta^2}} \right)}. \nonumber
\end{align}
Squaring on both sides and taking $\hbar v_F=1$, we obtain a quadratic equation $a E_n^2 + b E_n + c = 0$ with
\begin{align}
    a &= 1 \nonumber , \\
    b &= 2 \beta q_x - 3 \beta\left(1-\beta^2\right) |b n| ,\\
    c &= \beta^2 q_x^2 - 2(1-\beta^2)^{3/2} |b n| -  3  \left(1 - \beta^2\right) |b n| q_x . \nonumber
\end{align}
Neglecting terms of order $\mathcal{O}(b^2)$, we get
\begin{align}
   \frac{E_n}{\hbar v_F} \approx& -\beta q_x + \frac{3}{2} \beta (1-\beta^2) |b n| \\
   &+ {\rm sgn}(n) (1-\beta^2)^{3/4} \sqrt{|bn|(2 + 3 \sqrt{1-\beta^2} q_x)}. \nonumber
\end{align}
In the limit of small $\beta$ and small $q_x$ this expression can be further simplified by keeping terms up to second-order in any small quantity ($\beta, q_x, b$), leading to
\begin{align}
   \frac{E_n}{\hbar v_F} \approx -\beta q_x + \frac{3}{2} \beta |b n| + {\rm sgn}(n) (1-\beta^2)^{3/4} \sqrt{|bn| ( 2 +  3 q_x)},
\end{align}
which is Eq.~(\ref{eq:spectrum_b_e}) in the main text.

\end{document}